\documentclass[final,3p,times,twocolumn]{elsarticle}
\usepackage{amssymb}
\usepackage{lineno}
\usepackage[version=3]{mhchem}
\usepackage{gensymb}
\usepackage{textcomp}
\usepackage{xcolor}

\newcommand{\dtmin}{$D^2_\mathrm{min}$}
\newcommand{\tg}{$T_\mathrm{g}$}

\journal{}
\begin{document}
\begin{frontmatter}

\title{Large-scale atomistic study of plasticity in amorphous gallium oxide with a machine-learning potential}

\author[,1,3]{Jiahui Zhang} % \thanks{Corresponding author: jiahui.zhang@helsinki.fi}}%\corref{cor1}}
\ead{jiahui.zhang@helsinki.fi}
\author[2]{Junlei Zhao}
\author[1]{Jesper Byggmästar}
\author[3]{Erkka J. Frankberg}
\author[1]{Antti Kuronen}

\affiliation[1]{Department of Physics, University of Helsinki, P.O. Box 43, FI-00014, Helsinki, Finland}
\affiliation[2]{Department of Electrical and Electronic Engineering, Southern University of Science and Technology, Shenzhen, 518055, China}
\affiliation[3]{Materials Science and Environmental Engineering Unit, Tampere University, Tampere, 33720, Finland}

\cortext[cor1]{Corresponding author}

\begin{abstract}
Compared to the widely investigated crystalline polymorphs of gallium oxide (\ce{Ga2O3}), knowledge about its amorphous state is still limited. With the help of a machine-learning interatomic potential, we conducted large-scale atomistic simulations to investigate the glass transition and mechanical behavior of amorphous \ce{Ga2O3} (a-\ce{Ga2O3}). During the quenching simulations, amorphization of gallium oxide melt is observed at ultrahigh cooling rates, including a distinct glass transition. The final densities at room temperature have up to 4\% variance compared to experiments. The glass transition temperature is evaluated to range from 1234 K to 1348 K at different cooling rates. Structural analysis of the amorphous structure shows evident similarities in structural properties between a-\ce{Ga2O3} and amorphous alumina (a-\ce{Al2O3}), such as radial distribution function, coordination distribution, and bond angle distribution. An amorphous gallium oxide structure that contains approximately one million atoms is prepared for the tension simulation. A highly plastic behavior is observed at room temperature in the tension simulations, comparable to amorphous alumina. With quantitative characterization methods, we show that a-\ce{Ga2O3} can possibly has a higher nucleation rate of localized plastic strain events compared to a-\ce{Al2O3}, which can increase the material's resistance to shear banding formation during deformation.

\end{abstract}

\begin{highlights}
\item Verified the stability and reliability of the machine-learning potential of gallium oxide for the amorphous phase
\item Estimated the glass transition conditions for gallium oxide
\item Predicted the mechanical behaviors for amorphous gallium oxide through atomistic simulations
\end{highlights}


\end{frontmatter}

\section{Introduction}

Oxide glass materials are widely used in industry and daily life because of their diverse functionalities~\cite{idota1997, Kim2006}. However, one of the biggest limitation of their wider usage is their incapability in load bearing, despite of their extraordinary theoretical strength-to-density ratio. This is because these materials are generally considered brittle and flaw sensitive at room temperature due to the lack of an effective plastic deformation mechanisms. For example, unlike crystalline materials, the dislocation mediated deformation mechanisms do not exist in oxide glass materials. Nevertheless, recent discoveries have shown that some oxide glasses exhibit plasticity via diffusion based mechanisms. Amorphous silica is generally known as a brittle material, but computational studies have reported that it can show plasticity at room temperature if prepared under certain non-equilibrium conditions such as under high hydrostatic pressure or by ultra-fast cooling rate~\cite{Yuan2014, Lane2015}. Additionally, Frankberg \textit{et al.} reported that thin films of amorphous alumina (a-\ce{Al2O3}) exhibit plastic behavior under shear and compression as well as in tension at room temperature. Combined with computational results, they confirmed that time-dependent viscous creep is the dominant plastic deformation form and proposed an atomistic bond-switching mechanism that plays an essential role mediating the plastic deformation~\cite{Frankberg2019}. Microscopically, the bond switching occurs by the nucleation of localized plastic strain events (LPSEs), and at lower strain rates the plasticity is distributed more homogeneously in the volume, while at higher strain rates the plasticity is constrained to form shear bands~\cite{Frankberg2023}. Large-scale computational characterization methods are especially important for investigating plastic deformation mechanisms in amorphous materials, because small-scale methods suitable for crystalline materials, such as \textit{ab initio} calculations, can hardly be used for non-crystalline materials due to their lack of structural periodicity. Moreover, as the lifetime of a single LPSE is some nanoseconds and includes a small amount of atoms at a time, this is usually beyond the characterization capability of short-range structural analyzing methods. Therefore, to overcome this challenge, different computational characterization methods have been proposed to provide a quantitative indication of the plasticity in amorphous materials, such as topological constraint theory~\cite{Phillips1981, Thorpe1983}, \dtmin~analysis~\cite{Falk1998}, ring statistics~\cite{Gaskell1997, Du2004}, and coarse-grained analysis~\cite{Zhang2023}. Each of them come with advantages and disadvantages in quantitatively characterizing the plasticity and disclosing the medium-range structural information and combining different methods make the formidable challenge of quantifying and comparing plasticity across different oxide glass materials more feasible.

Beyond the existing studies of low-temperature plasticity in amorphous silica and alumina, the current body of literature offers limited coverage on other diatomic oxide glass materials. This study is dedicated to characterize \ce{Ga2O3} which has analogous crystalline phases with \ce{Al2O3}~\cite{Levin1998}. In addition, Ga$^{3+}$ and Al$^{3+}$ cations have distinct chemical similarity and outer electron structure and oxides of Ga and Al form solid solutions together without phase changes~\cite{HILL1952}. Due to this similarity, these two materials are often investigated together and compared with each other~\cite{wang_band_2018, Han2020, Mu2022}. Among the known crystalline phases, the $\alpha$, $\beta$, $\gamma$, $\delta$, and $\kappa$ phases are the (meta-)stable and clearly distinguishable ones, apart from amorphous \ce{Ga2O3} (a-\ce{Ga2O3}). Crystalline gallia is a promising semiconductor material because of its ultrawide bandgap ($\sim$4.85-5.35 eV)~\cite{Tadjer2022}, while a-\ce{Ga2O3} is a novel functional material, candidate for photodetectors~\cite{Cui2017, Liang2019, Qin2019}. Experimental studies on crystalline \ce{Ga2O3} have become increasingly popular in recent years, but a-\ce{Ga2O3} has been studied only at a limited level. As a poor glass former, again similar to \ce{Al2O3}, the synthesis of a-\ce{Ga2O3} requires non-equilibrium conditions, such as ultrahigh cooling rates, which limits the available synthesis methods and size of the sample that can be obtained with the current technology.

From a computational perspective, density functional theory (DFT) and \textit{ab initio} molecular dynamics (AIMD) methods have been extensively used in investigating \ce{Ga2O3} and can provide data of high accuracy~\cite{Schubert2016, Furthmuller2016, ponce2020, Mu2022, Lion2022}. However, the high computing power cost of DFT and AIMD methods makes a larger-scale computational study of \ce{Ga2O3} impossible. Characterizing plasticity especially benefits from large-scale atom systems in order to avoid the occurrence of finite-size effects, as been observed when investigating other amorphous oxide materials~\cite{Yuan2014, Zhang2020}. Therefore molecular dynamics (MD) studies at a larger scale (more than 10$^{4}$ atoms) have become of great importance in studying plasticity in amorphous oxides \cite{tang_brittle--ductile_2023, Wu2021}.

Machine-learning (ML) methods are emerging as important and efficient tools in interatomic potential (IAP) development~\cite{Unke2021}. From a large enough DFT calculation database, it is possible to obtain an IAP that has accuracy comparable to \textit{ab initio} methods and is highly efficient to compute at the same time. IAPs for specific crystalline phases of \ce{Ga2O3} have already been developed~\cite{Liu2020, liu2023}. Recently, a low-dimensional tabulated Gaussian approximation potential (tabGAP) aimed to provide a solution for the universal atomistic studies of \ce{Ga2O3} was developed \cite{Zhao2023}. With the help of this ML-IAP, it is now possible to extend the computational studies on amorphous \ce{Ga2O3} to larger length and time scales, avoiding finite-size effects when investigating the plastic deformation behavior. MD simulations with the ML-IAP can provide an accurate prediction regarding the mechanical behavior and mechanisms of plasticity in amorphous \ce{Ga2O3}, providing a deeper understanding of the fundamental mechanisms and facilitating future experimental studies to validate these results.

In this work, we computationally prepared a-\ce{Ga2O3} with the ML-IAP through a melt-quenching scheme. The obtained structure was characterized and compared with existing literature on amorphous \ce{Ga2O3}. Next, tensile test simulations were conducted on large-scale amorphous structure including approximately one million atoms to investigate the mechanical behavior and plasticity of a-\ce{Ga2O3}. Plasticity observed under tension was quantitatively characterized and compared with earlier experimental and computational results on a-\ce{Ga2O3} and on a-\ce{Al2O3}, the latter being a well-known amorphous oxide capable of significant ductility at room temperature~\cite{Frankberg2019}. We show that a-\ce{Ga2O3} shows similar low temperature plasticity as a-\ce{Al2O3} suggesting that high ductility is not unique to a-\ce{Al2O3} but can be generalized to other amorphous oxides that fulfill the criteria introduced here to be shared between a-\ce{Al2O3} and a-\ce{Ga2O3}.

\section{Methodology}

\subsection{Preparation of the Amorphous Structure}
% \subsection{Simulation Details}

Melt-quenching molecular dynamics (MD) simulations on \ce{Ga2O3} is performed under various cooling rates. To control the computing resource needed, a stoichiometric system with 48,000 randomly generated atoms is first heated up to 3000 K and equilibrated for 50 ps. Then the structure is cooled down to 300 K with desired cooling rate and then equilibrated for another 50 ps. Periodic boundary conditions and isothermal-isobaric ($NPT$) ensemble is applied to the system throughout the simulation. Temperature and pressure is controlled using the Nos{\'e}-Hoover algorithm~\cite{hoover1985}. Four independently prepared structures are used in the glass transition analysis.  All MD simulations are performed using the LAMMPS code~\cite{thompson_lammps_2022} with the recently developed general-purpose ML-IAP (tabGAP) for \ce{Ga2O3}~\cite{Zhao2023}.

\begin{figure}[htbp!]
\centering\includegraphics[width=\linewidth]{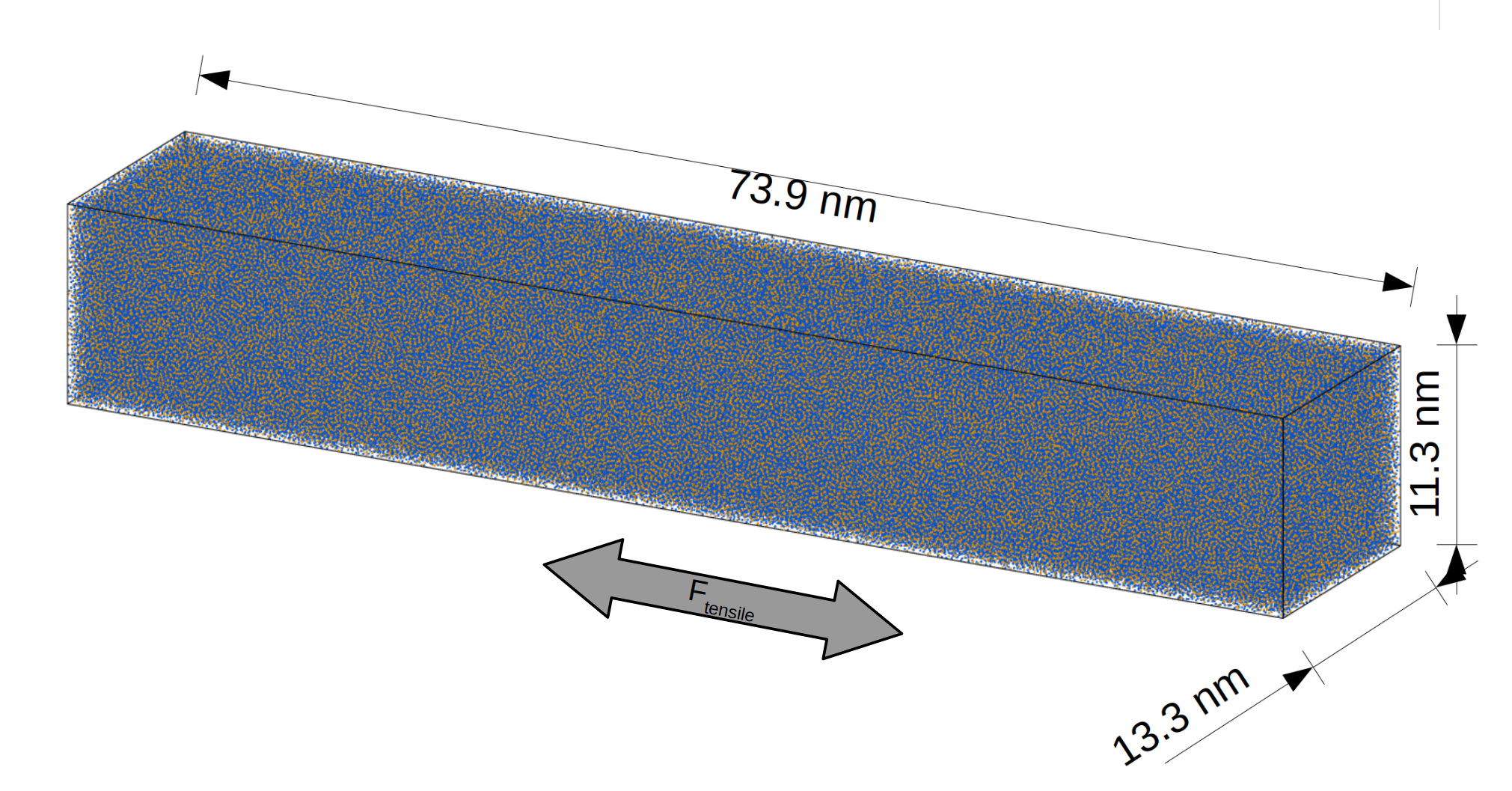}
\caption{Amorphous structure with 960,000 atoms after melt-quenching. Gallium and oxygen atoms are colored in yellow and blue, respectively. The gray arrow indicates the direction of the tensile force, $F_\mathrm{tensile}$.}
\label{fig:gao}
\end{figure}

Another a-\ce{Ga2O3} structure with significantly larger dimensions is prepared for the tensile test simulations using a cooling rate of $1\times10^{12}$ K/s. The final size of this structure is $11.3\times13.3\times73.9$ nm$^{3}$ with 960,000 atoms, as shown in Figure~\ref{fig:gao}.

To compare the mechanical behavior, an a-\ce{Al2O3} bulk structure with 960,000 atoms is created in a similar melt-quenching fashion as described in Ref.~\cite{Gutierrez2002}. A Buckingham-form IAP developed by Matsui is used for a-\ce{Al2O3} simulations~\cite{Matsui1994}.

\subsection{Tensile Test Settings}

During the tensile simulations, structures are deformed along the long axis. Periodic boundary conditions were applied to the system. The $NPT$ ensemble is applied on dimensions orthogonal to the tensile deformation, the temperature and pressure is kept at 300 K and 0 bar during the stretching using the Nos{\'e}-Hoover algorithm. The simulation box is deformed every MD timestep (1 fs) without remapping the atom coordinates. The strain rate is $5\times10^{8}$~s$^{-1}$, and the simulation box is stretched to a maximum of 50\% engineering strain. In this work, we use engineering strain and true stress in all results. The strain is defined as:
\begin{equation}
\varepsilon = \frac{L-L_{0}}{L_{0}},
\label{eq:strain}
\end{equation}
where $L_{0}$ is the original length of the system and $L$ is the length at a given strain. True stress is equivalent to the momentary tensile force divided by the cross-sectional area perpendicular to the tensile axis at the corresponding moment. Radial distribution function (RDF), bond angle distribution (BAD) and coordination distribution (CD) are analyzed using the OVITO analysis and visualization software~\cite{Ovito}. For CD and BAD analyses, a cutoff distance of 2.3~\r A is used for both a-\ce{Ga2O3} and a-\ce{Al2O3}, which is the first minimum in the RDFs.

\subsection{Characterization of Plasticity}

Based on the tensile simulation results, local plastic strain (\dtmin) is calculated in a-\ce{Ga2O3} and a-\ce{Al2O3}. \dtmin\ is the minimum of the non-affine squared displacement as introduced by Falk and Langer~\cite{Falk1998} and has been proven useful in capturing plastic deforming regions. In this work, we calculate momentary \dtmin, which means that the reference configuration changes with the configuration to be analyzed. A constant 1\% strain is used between these two configurations. The same cutoff distances used in \dtmin\ analysis are 3.0 \r A, 4.6 \r A, and 6.0 \r A.

Based on the \dtmin\ analysis, the concept of LPSEs is defined in Refs.~\cite{Frankberg2019, Frankberg2023}. Although the size of LPSEs can vary a lot, here we define the LPSE according to Ref.~\cite{Frankberg2023} to make a meaningful comparison to earlier results with a-\ce{Al2O3}. Therefore, LPSE refers here to an atom cluster that has more than 200 atoms that each have a \dtmin\ higher than the threshold value. To quantify the occurrence of these events, cluster analysis is performed on the atoms fulfilling the LPSE definition. The threshold value equals two times the average momentary \dtmin\ of the system and would dynamically change during the tensile test. The total amount of atoms included in the clusters and the size distribution of the clusters are then further compared.

The polyhedral analysis is performed in a way proposed by Zhang \textit{et al.}~\cite{Zhang2023}. Polyhedron in this work indicates a structure that has one center cation and multiple bonded oxygen atoms. According to the way they connect, polyhedra are classified as corner-sharing (CS), edge-sharing (ES), and face-sharing (FS).

\section{Results}
\subsection{Amorphization and Glass Transition of Gallium Oxide}

\begin{figure*}[ht!]
\centering\includegraphics[width=\linewidth]{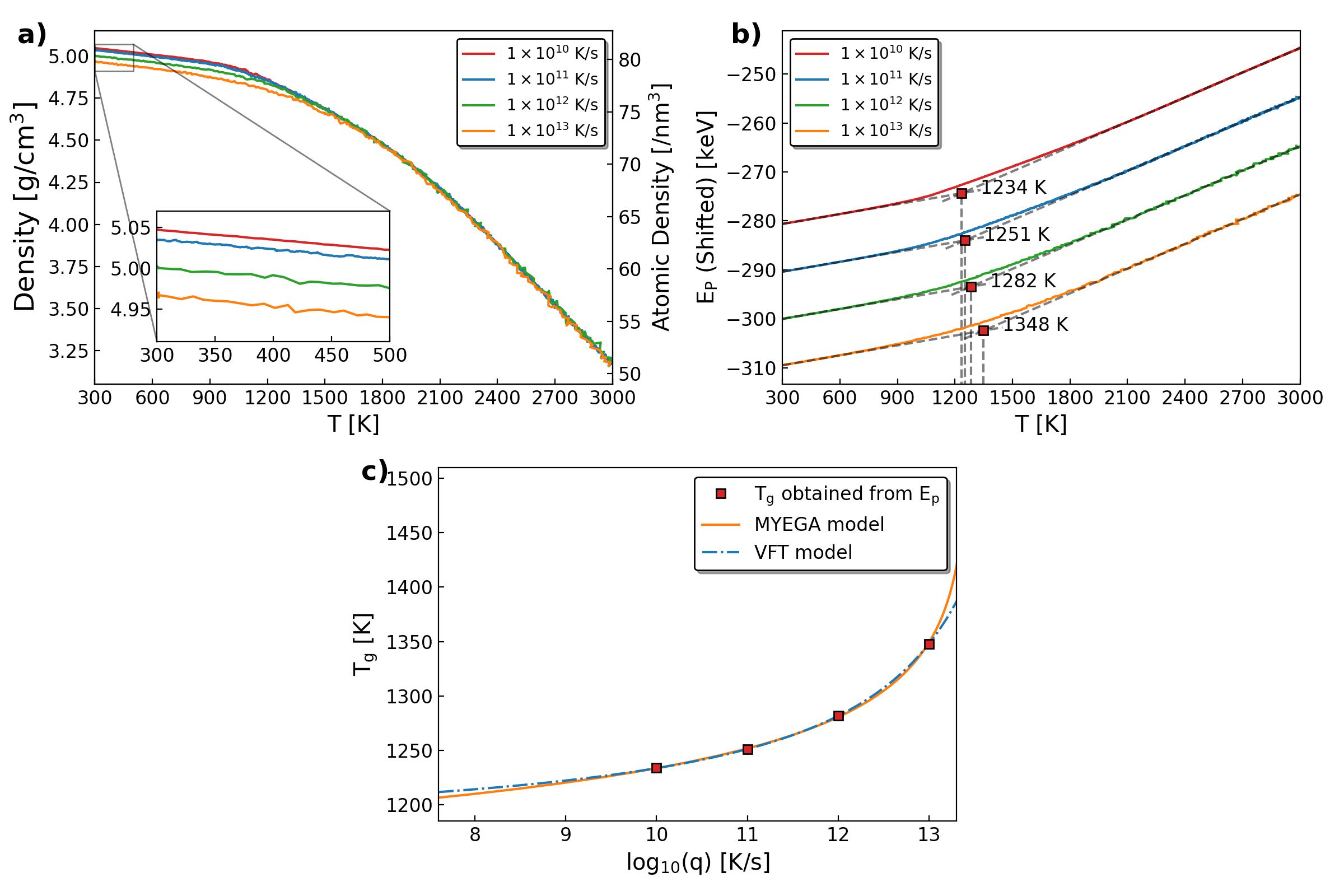}
\caption{(a) The mass density ($\rho$) and atomic density of the \ce{Ga2O3} structure as a function of temperature, with different cooling rates. (b) Potential energy ($E_\mathrm{p}$) as a function of temperature during cooling, with different cooling rates. (c) Fit of the $T_\mathrm{g}$ data evaluation from potential energy change to the VFT and the MYEGA model.}
\label{fig:gt}
\end{figure*}

The four cooling rates ($q$) used in the melt-quenching fashion preparation of a-\ce{Ga2O3} structures were $1\times10^{10}$ K/s, $1\times10^{11}$ K/s, $1\times10^{12}$ K/s, and $1\times10^{13}$ K/s. Figure~\ref{fig:gt}(a) presents the mass density ($\rho$) and atomic density of a-\ce{Ga2O3} as functions of temperature during quenching. At 3000 K, liquid \ce{Ga2O3} has a density of 3.2 g/cm$^3$. Above 1800 K, curves of density under different cooling rates overlap with each other, showing a linear increase with decreasing temperature. As temperature decreases below 1800 K, the increase of density slows down and the curves start to diverge from each other, leading to differences at the final temperature of 300 K. The highest cooling rate produced a density of 4.96 g/cm$^3$, and the lowest cooling rate 5.05 g/cm$^3$, with a 2\% difference. At 300 K, the a-\ce{Ga2O3} densities show approximately a 4\% difference with recent experimental results of 4.78 to 4.84 g/cm$^3$~\cite{Han2020,liu2023}, consistent with the density range reported earlier~\cite{Yu2003}. However, we also report a different density of approximately 4.25 g/cm$^3$ at 2100 K, which is lower than the reported experimental value of 4.84 g/cm$^3$~\cite{Dingwell1992}.

Regarding the atomic density, at the chosen cooling rates we obtain atomic densities ranging from 79 to 81 atoms per nm$^{3}$ at 300 K. Results show that though a-\ce{Ga2O3} has a significantly higher mass density due to heavier Ga atoms, its structure is not as closely packed as a-\ce{Al2O3}, which has an atomic density of approximately 97 atoms per nm$^{3}$ at 300 K~\cite{Zhang2024}.

The potential energy ($E_\mathrm{p}$) as a function of temperature for all cooling rates is presented in Figure~\ref{fig:gt}(b). At all cooling rates, potential energy shows a continuous and smooth decrease without abrupt change, indicating a transition from liquid to amorphous state instead of crystallization. The evident difference of gradients at high- and low-temperature parts of the curves indicates possible glass transition during the quenching. We determined the glass transition temperature (\tg) by separately fitting the high- and low-temperature parts of the $E_\mathrm{p}$-$T$ curve using a linear function. \tg\ is determined as the intersection point of the two fitted lines ranging between 1234 to 1348 K and showing high dependence on the cooling rate. 

% The values obtained are listed in Table~\ref{tab:Tg}.

% \begin{table}[]
%     \centering
%     \caption{Glass transition point (\tg) of \ce{Ga2O3} estimated using the potential energy data of melt quenched systems produced using different cooling rates.}
%     \begin{tabular}{c c}

%     \hline
%          $q$ (K/s) & $T_\mathrm{g} (K)$ \\
%     \hline
%          $1\times 10^{10}$ &  1234 \\
%          $1\times 10^{11}$ &  1251 \\
%          $1\times 10^{12}$ &  1282 \\
%          $1\times 10^{13}$ &  1348 \\
%     \hline
%     \end{tabular}
%     \label{tab:Tg}
% \end{table}

A number of models have been proposed to describe the correlation between viscosity and temperature of glass-forming materials, from which the correlation between glass transition point and cooling rate can possibly be derived. Among them, the Vogel–Fulcher–Tammann (VFT) model is an early and widely used one, but is also believed to behave poorly for fragile glass formers~\cite{vogel_temperaturabhangigkeitsgesetz_1921, fulcher_analysis_1925, tammann_abhangigkeit_1926, Angell2000}. The Mauro-Yue-Ellison-Gupta-Allan (MYEGA) model is a newer model that has been shown to behave better for poor glass formers~\cite{Mauro2009}. The correlation between \tg\ and $q$ is based on Maxwell's expression between relaxation time and viscosity:
\begin{equation}
    \tau = \frac{\eta}{G_{\infty}},
\end{equation}
where $G_{\infty}$ is the instantaneous shear modulus, and the assumption of an inverse correlation between the relaxation time and cooling rate:
\begin{equation}
    \tau = \frac{T_{1}}{q} \bigg|_{T=T_\mathrm{g}},
\end{equation}
where $T_{1}$ is a constant. Least-squares method curve fitting is performed with the following equations and results are shown in Figure~\ref{fig:gt}(c). For the VFT model, the fitting function is:
\begin{equation}
    T_\mathrm{g} = T_{0} - \frac{A}{\log_{10}(Bq)},
\end{equation}
where $A$, $B$ and $T_0$ are fitting parameters. From our data, we obtain $A=305.636$ K, $B=1.9679\times10^{-15}$ s/K, $T_{0} = 1168.87$ K. For the MYEGA model, the fitting function is:
\begin{equation}
    \log_{10}(q) = A - \frac{B}{T_\mathrm{g}} \exp(\frac{C}{T_\mathrm{g}}),
\end{equation}
where $A$, $B$ and $C$ are the fitting parameters. And here we obtain $A=13.4607$ K/s, $B=7.1997\times10^{-7}$ K$^2$/s, $C=2.7763\times10^4$ K. We see from Figure~\ref{fig:gt}(c) that both models can give a smooth fit to the \tg\ values obtained from our simulations. The MYEGA model predicts slightly higher \tg\ at $q>10^{13}$ K/s, and lower prediction of \tg\ at $q<10^{10}$ K/s, compared to the VFT model. The steeper increase of predicted \tg\ at high $q$ region is reasonable if we consider the significant artifacts that have been observed in amorphous silica when prepared at a $q=10^{14}$ K/s~\cite{Yuan2012}. It is possible that the behavior of the structure detaches quickly from a predictable behavior beyond $q>10^{14}$ K/s and such cooling rates remain out of experimental reach. At the low $q$ region, verifying the performance of the model is challenging because of the occurrence of crystallization. For a-\ce{Al2O3}, crystallization has been observed at a computational cooling rate of $10^{10}$ K/s~\cite{Zhang2024}. Although not observed in this computational work, a-\ce{Ga2O3} is also assumed to require extreme cooling rates to remain amorphous, due to its reported strong tendency for crystallization~\cite{Zhao2023, SFzhao2024crystallization}.

\subsection{Tensile Test}

\begin{figure*}[h!]
\centering\includegraphics[width=\linewidth]{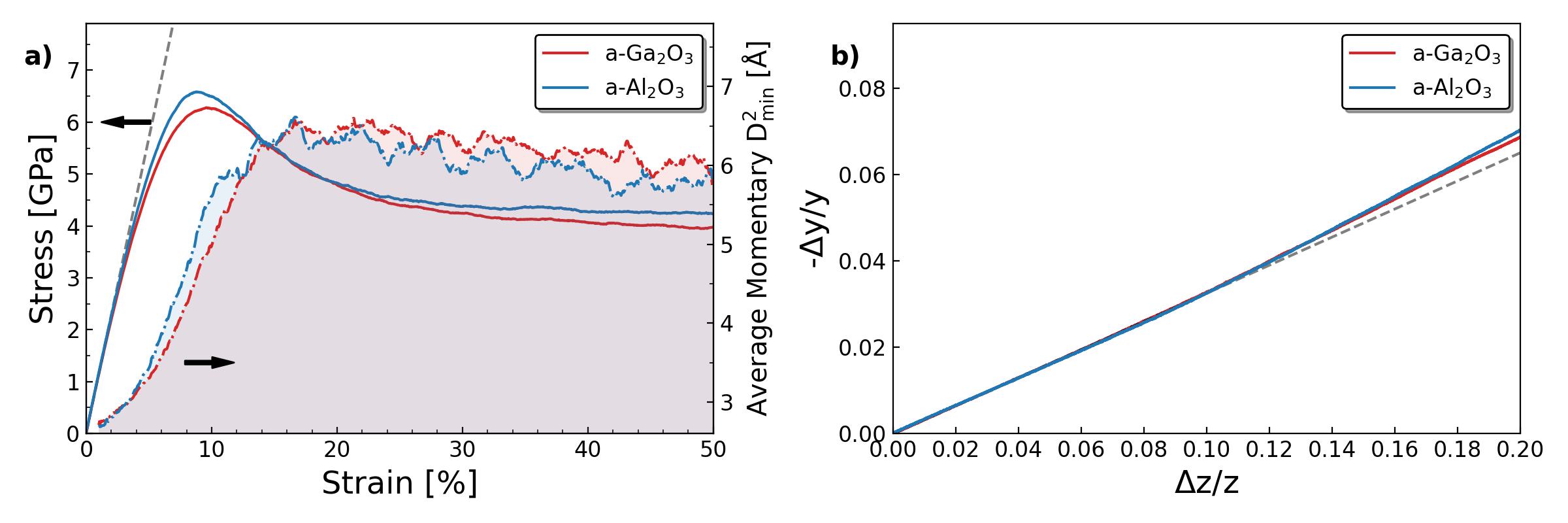}
\caption{(a) Stress (solid lines) and average momentary (\dtmin\, dashed lines) as functions of strain during tensile test. The Young's modulus is calculated from the linear fitting of the first 1\% of the curve. The \dtmin\ is calculated with $r_\mathrm{cut}=4.6$ \r A. (b) Volume change on the dimension orthogonal to the strained dimension as a function of the volume change on the strained dimension in both a-\ce{Ga2O3} and a-\ce{Al2O3}. The Poisson's ratio of a-\ce{Ga2O3} is fitted from the first 1\% strain ($\Delta z/z <0.01$) data of a-\ce{Ga2O3}.}
\label{fig:stress}
\end{figure*}

Tensile test simulations are performed using a large-scale amorphous structure prepared with $q=1\times10^{12}$ K/s to avoid possible finite-size effects. The structure is elongated to 50\% engineering strain at a strain rate of $5\times10^{8}$ s$^{-1}$. In Figure~\ref{fig:stress}(a), we present the stress-strain results, including data of a-\ce{Al2O3} obtained under the same conditions. For both structures, the tensile stress first increases linearly with strain until the yielding. From the elastic deformation region, the Young's modulus of a-\ce{Ga2O3} is calculated by a linear fitting of the first 1\% of the stress-strain curve. For a-\ce{Ga2O3}, we obtain the Young's modulus of $113.74\pm0.22$ GPa. As a comparison, the elastic modulus evaluated from nanoindentation experiments ranges from 100 to 220 GPa~\cite{battu2018}. The yielding occurs later in both structures and they reach maximum stresses of 6.2 GPa for a-\ce{Ga2O3} and 6.6 GPa for a-\ce{Al2O3}. After yielding, the stress gradually decreases and levels to a steady state flow stresses of 3.9 GPa for a-\ce{Ga2O3} and 4.2 GPa for a-\ce{Al2O3} measured at 50\% strain. While a-\ce{Ga2O3} has a slightly lower strength, it also shows further minor softening during the steady state flow up to 50\% strain.

Figure~\ref{fig:stress}(a) also presents the average momentary \dtmin\ measured during the tensile tests. Note that the \dtmin\ results between a-\ce{Ga2O3} and a-\ce{Al2O3} are comparable here because we are able to use the same cutoff distance ($r_\mathrm{cut}$) when performing the \dtmin\ analysis, which will be discussed later. In comparison, we see that these two materials have similar average \dtmin\ results. Atoms in a-\ce{Al2O3} response to strain slightly faster than a-\ce{Ga2O3} as indicated by the momentary \dtmin\ value increase at a range of 2-10\% strain, while a-\ce{Ga2O3} shows a slightly higher average during the steady state flow at range of 20-50\% strain. No other significant difference in plasticity can be confirmed from the average momentary \dtmin\ results. To obtain a quantitative comparison of the plastic deforming ability between these two materials, further characterization is needed.

Figure~\ref{fig:stress}(b) shows the correlation between relative deformation in directions parallel ($Z$-axis) and perpendicular ($Y$-axis) to the tensile force. Similar to the stress-strain curve, a correlation with a linear function (dashed line) between the length changes can be observed up to 10\% engineering strain, while after 10\% the values start to diverge from linearity. The Poisson's ratios ($\nu$) are calculated from the 1\% strain of the data. The result is 0.3253 for a-\ce{Ga2O3}, and 0.3215 for a-\ce{Al2O3}. The measured Poisson's ratio corresponds approximately to that of metallic aluminum alloys and such high values of the ratio have been connected to increased shear deformation ability of amorphous oxide materials~\cite{Garcia2013, Frankberg2019}. In addition, Poisson's ratio has been reported to have the following correlation with a glass former's fragility, which can be used to classify liquids into strong or fragile glass formers~\cite{novikov_poissons_2004}:
\begin{equation}
    m - 17 = 29(\frac{B_{0}}{G} - 1),
\end{equation}
where $B_{0}$ is the bulk modulus, $G$ is the shear modulus, and $m$ is fragility. The correlation between $B_{0}$, $G$, and $\nu$ is formulated as:
\begin{equation}
    \frac{B_{0}}{G} = \frac{2}{3} \frac{(1+\nu)}{(1-2\nu)},
\end{equation}
thus the value of $B_{0}/G$ can be calculated to be 2.5287 for a-\ce{Ga2O3}, and 2.4678 for a-\ce{Al2O3}. The fragility is then calculated as 61.3328 for a-\ce{Ga2O3} and 59.5658 for a-\ce{Al2O3}. Direct estimation of fragility is difficult because the high viscosity at temperatures near \tg\ makes the time cost unaffordable. Therefore, from the data in a much higher temperature range, the fragility of a-\ce{Al2O3} has been estimated by fitting the super-Arrhenius response in the Angell plot~\cite{Wilding2010}, leading to the result of approximately 40. Therefore, both a-\ce{Ga2O3} and a-\ce{Al2O3} can be classified as fragile glass formers based on the simulation data.

\subsection{Structural Analysis}

\begin{figure*}[ht!]
\centering\includegraphics[width=\linewidth]{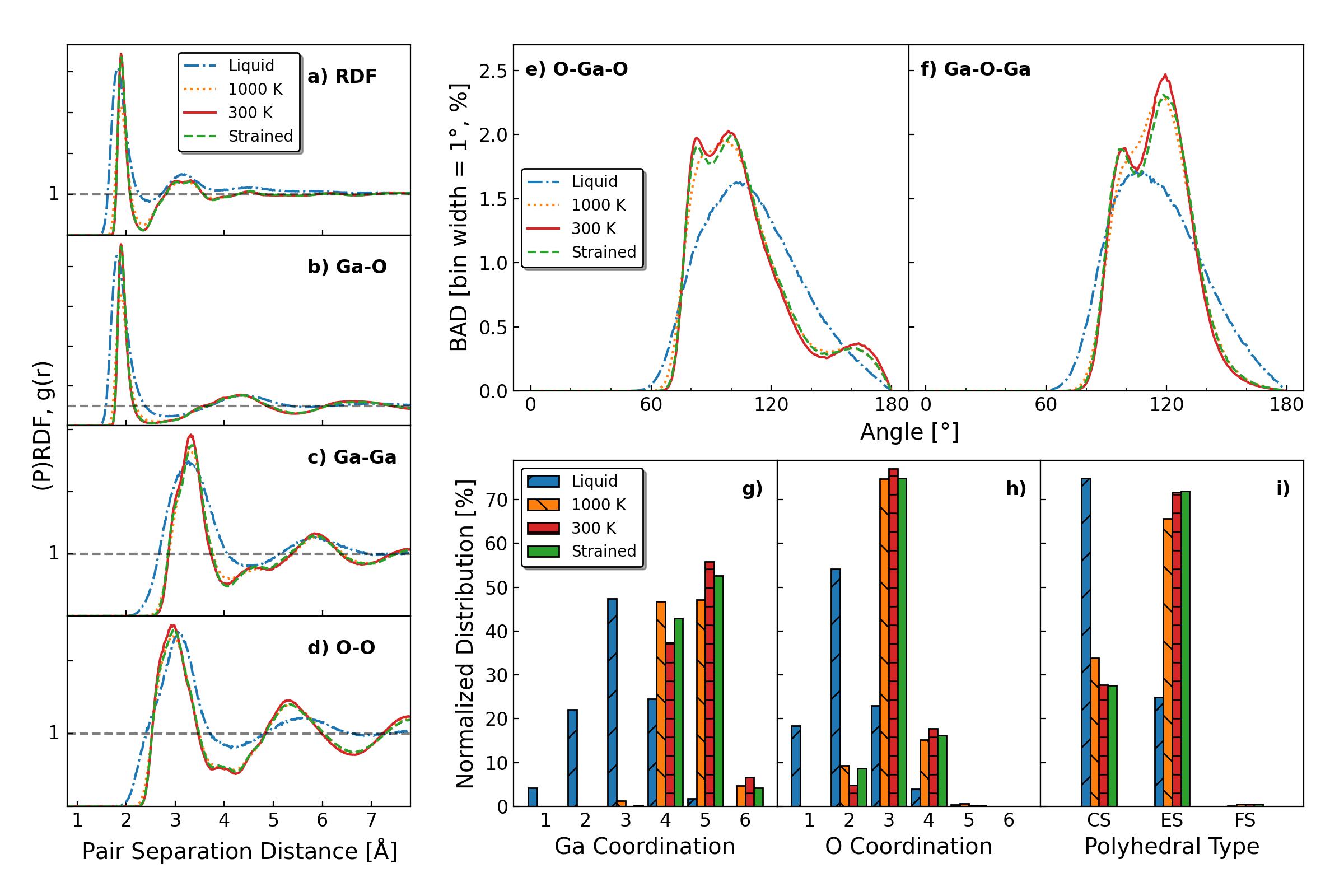}
\caption{(a-d) Radial distribution function and partial radial distribution functions of the a-\ce{Ga2O3} structures. (e, f) Bond angle distribution functions of O-Ga-O (Ga centered) and Ga-O-Ga (O centered) bond angles at 3000 K, 1000 K, 300 K and 50\% strained conditions. Cutoff distance is 2.3 \r A for creation of bonds.The bin size is $1\degree$ and the results are normalized by the total bond number. (g, h) Coordination distributions of Ga atoms, O atoms, and (i) polyhedra type at 3000 K, 1000 K, 300 K, and 50\% strained conditions. The results are normalized by the total number of the corresponding type.}
\label{fig:analysis}
\end{figure*}

After obtaining the strained structure, a complete characterization is performed to verify that the structure prepared for tensile test simulation is amorphous, and investigate the influence of strain on the structural features. We compute the RDF and BAD results for the four conditions: liquid (3000 K), near the \tg\ (1000 K), room temperature (unstrained, 300 K), and 50\% strain (strained, 300 K). Results are shown in Figure~\ref{fig:analysis}(a-f). The RDF results show that for all structures there is a clear convergence to $g(r)=1$ as the pair distance increases, confirming the absence of ordering in the long range. The RDF and pairwise partial RDFs (PRDFs) of liquid \ce{Ga2O3} show evident differences from the other three structures [Figure~\ref{fig:analysis}(a-d)], and much wider peaks can also be observed in the BAD plots [Figure~\ref{fig:analysis}(e, f)]. This is due to the low density and high mobility of the system at high temperature. In comparison, from 1000 K to 300 K, only minor differences are observed, which confirms the occurrence of glass transition above this temperature range between 1000 and 3000 K. At 300 K, the first g(r) peak values for Ga-O, Ga-Ga, O-O pairs appear at 1.90 \r A, 3.31 \r A, and 2.95 \r A, respectively. Although the structure is quenched without fixing the volume as constant, the RDF results are closely related to the DFT results reported earlier~\cite{liu2023}. Comparing unstrained and strained results at 300 K, we see that strain has only a minor influence on the RDFs and PRDFs.

The BADs of the O-Ga-O and Ga-O-Ga triplets are presented in Figure~\ref{fig:analysis}(e-f). For liquid \ce{Ga2O3}, both the O-Ga-O and Ga-O-Ga bond angles have a broad distribution with a single peak value around $100\degree$. From 1000 K to 300 K, the overall bond angle distribution show no change in position, but clear peaks can be seen to form at approximately $80\degree$, $100\degree$, and $170\degree$ for O-Ga-O and $100\degree$, and $120\degree$ for Ga-O-Ga indicating an increased directionality of the bonding.

Figure~\ref{fig:analysis}(g, h) presents the CD of gallium and oxygen atoms and Figure~\ref{fig:analysis}(i) shows the polyhedral analysis results. Similarly to the RDF and BAD results, the most significant difference can be observed in the 3000 K liquid structure, where much lower coordination numbers (such as 1-fold and 2-fold) of gallium and oxygen atoms can be observed. We see that at 300 K, 56\% of the gallium atoms are 5-fold coordinated, 38\% are 4-fold coordinated and 6\% are 6-fold coordinated and straining at 300K has only a minor effect shifting atoms slightly from 5-fold and 6-fold coordination to 4-fold coordination. For oxygen atoms, 77\% of them are 3-fold coordinated, followed by 18\% 4-fold coordinated and 5\% 2-fold coordinated. From 1000 K to 300 K and then to strained structures, the coordination distributions show only minor differences. Polyhedral analysis results in Figure~\ref{fig:analysis}(i) show that in 300 K unstrained a-\ce{Ga2O3}, the ES polyhedra outnumber the CS polyhedra by an order of 2.7 to 1. The high fraction of ES polyhedra has been found to be correlated with the plastic deformation ability of amorphous oxide materials~\cite{Zhang2023}, which partially explains why a-\ce{Ga2O3} exhibits high ductility.

\begin{figure*}[ht!]
\centering
\includegraphics[width=\linewidth]{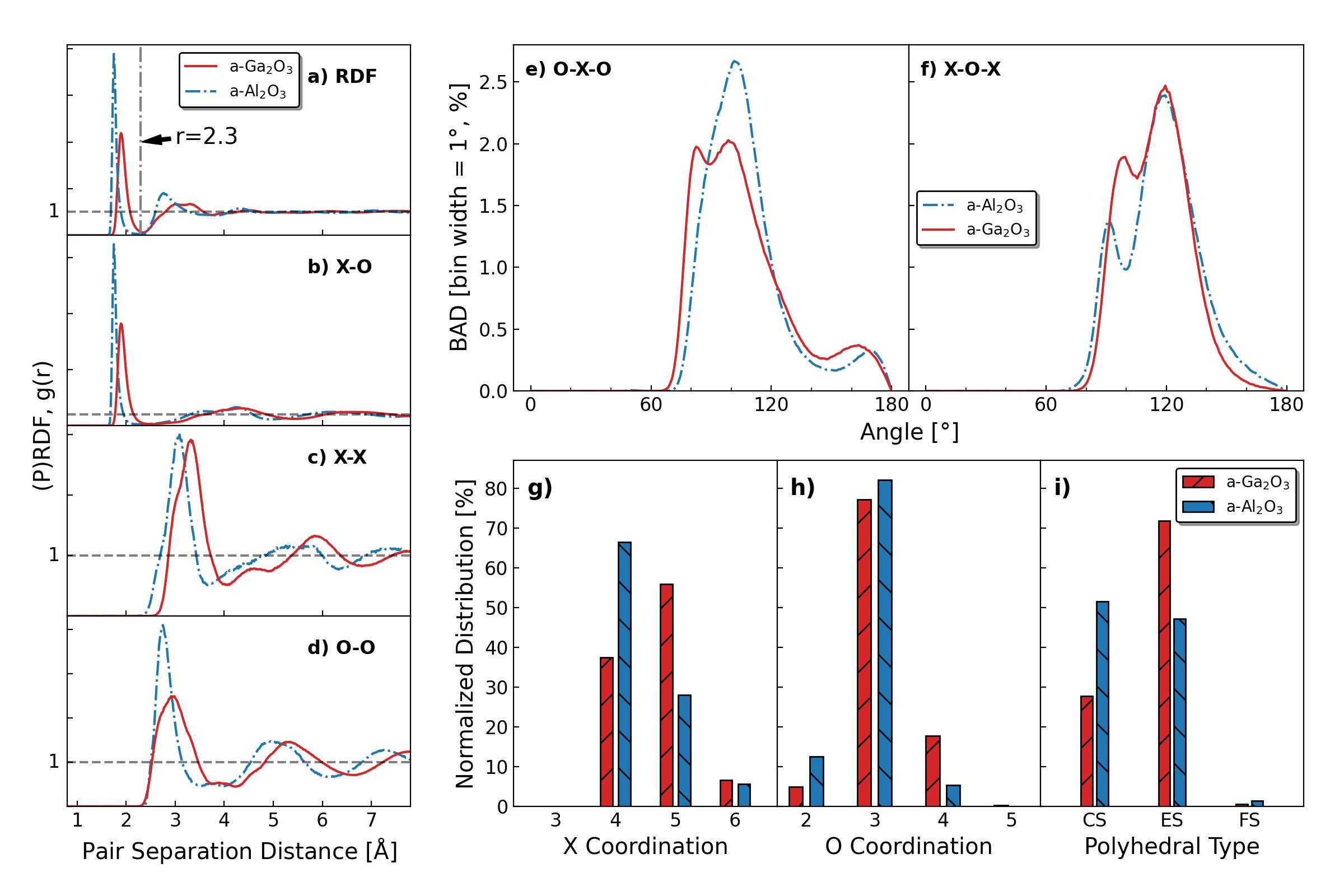}
\caption{(a-d) RDF and PRDFs of the a-\ce{Ga2O3} and a-\ce{Al2O3} structures at 300K. (e, f) Bond angle distribution of a-\ce{Ga2O3} and a-\ce{Al2O3} structure. X indicates cation atom Ga or Al. Cutoff distance is 2.3 \r A in creation of bonds.The bin size is $1\degree$ and the results are normalized by the total number of bonds. (g, h) CDs of cations and O atoms, and (i) polyhedra type in a-\ce{Ga2O3} and a-\ce{Al2O3}. Results are normalized by the total number of the corresponding type.}
\label{fig:compare}
\end{figure*}

Comparisons of a-\ce{Ga2O3} and a-\ce{Al2O3} are presented in Figure~\ref{fig:compare}. Specifically, in Figure~\ref{fig:compare}(a), we see that choosing 2.3 \r A as the cutoff distance for the \dtmin\ analysis is reasonable because it is indeed where the first minimum is for both the a-\ce{Ga2O3} and the a-\ce{Al2O3}. However, the first peak position of the Ga-O pair is found at 1.90 \r A, which is 0.15 \r A larger than the first peak of Al-O pairs, indicating different cation-oxygen bond length in the two materials. This consequently leads to the lower atomic density of a-\ce{Ga2O3} shown earlier. The first peak of a-\ce{Ga2O3} RDF is also significantly wider in comparison, which indicates that more variations of bond length are permitted within the structure. A similar trend can be seen in the PRDFs in Figure~\ref{fig:compare}(b-d) as well. BAD results in Figure~\ref{fig:compare}(e, f) show that both types of bond angles of a-\ce{Al2O3} are larger than that of a-\ce{Ga2O3} in general. O-Al-O has only one peak value at $100\degree$ instead of two present in O-Ga-O, and the first peak of the Al-O-Al distribution is slightly smaller and lower in height compared to Ga-O-Ga. Data on atomic coordination in Figure~\ref{fig:compare}(g, h) show that a-\ce{Ga2O3} has a significantly higher fraction of 5-fold coordinated and a significantly lower fraction of 4-fold coordinated Ga atoms compared to a-\ce{Al2O3}. Higher coordination of Ga atoms is likely enabled by the wider distribution of allowed bond lengths shown in Figure \ref{fig:compare}(a, b). The 3-fold coordinated O atoms are similar in number, but there are also more 4-fold coordinated O atoms and fewer 2-fold coordinated O atoms in a-\ce{Ga2O3}. Figure~\ref{fig:compare}(i) shows the medium-range polyhedral ordering, where a-\ce{Ga2O3} has a significantly higher fraction of ES polyhedra than a-\ce{Al2O3}, indicating a good potential for low temperature plasticity in a-\ce{Ga2O3}.

\subsection{Characterization of plasticity}

\begin{figure*}[ht!]
\centering\includegraphics[width=\linewidth]{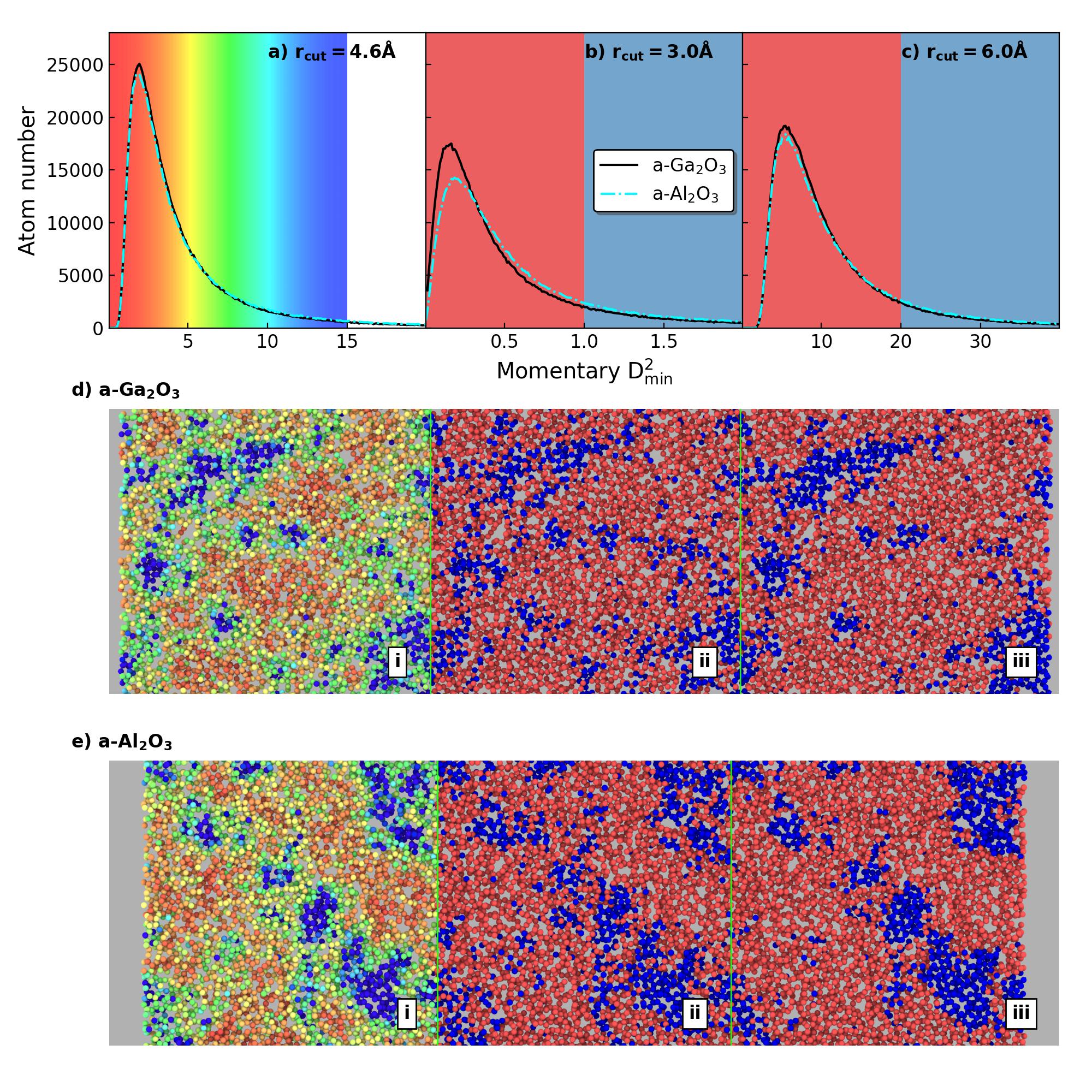}
\caption{Momentary \dtmin\ characterization of a-\ce{Ga2O3} and a-\ce{Al2O3} at 50\% strain. (a-c) Momentary \dtmin\ distributions calculated using different $r_\mathrm{cut}$. Background color indicates the color coding methods for panels (d, e). (d, e) Visualization of \dtmin calculated using different $r_\mathrm{cut}$ values. The visualized cross-section is taken along the long edge of the structures. The thickness of the slab is 5 \r A for both structures, and the width is approximately 11 \r A for a-\ce{Ga2O3} and 10 \r A for a-\ce{Al2O3}. Atoms are colored by their momentary \dtmin value, as indicated by the background color gradient of (a-c).}
\label{fig:d2min}
\end{figure*}

The stress-strain results in Figure~\ref{fig:stress}(a) show that the average \dtmin\ results are similar for a-\ce{Ga2O3} and a-\ce{Al2O3} from 1\% to 50\% strain, possibly indicating a similar plastic deforming ability. However, the average \dtmin\ is mainly a statistical quantity. Therefore, to reveal how similar the plastic behavior is between a-\ce{Ga2O3} and a-\ce{Al2O3}, we further analyze the \dtmin\ results from a microscopic point-of-view. When computing the momentary \dtmin, a larger cutoff would produce a generally larger value of \dtmin, and it helps to capture the localized deformation in a longer range and filter out the noise of short-range activities such as thermal vibration. On the other hand, summation within a large cutoff means that the smaller localized deformations are concealed. In contrast, a small cutoff distance can show localized deformation with a smaller size, but it would also produce more noise in the data. Combining these different results, we can identify a clearer spatial distribution of the localized deformation regions in these two materials. To probe the size of the localized deformation region in these materials, we calculate the momentary \dtmin\ of both structures at 50\% strain with different cutoff values. In Figure~\ref{fig:d2min}(a-c), we present the momentary \dtmin\ distribution with $r_\mathrm{cut}=4.6$ \r A, $r_\mathrm{cut}=3.0$ \r A, and $r_\mathrm{cut}=6.0$ \r A. Choosing 4.6 \r A is based on the two times the first RDF minimum principle, while 3.0 \r A and 6.0 \r A are used for comparison between a-\ce{Ga2O3} and a-\ce{Al2O3}. We see that although $r_\mathrm{cut}$ is decisive on the absolute value of \dtmin\ and on the amount of atoms associated with the maximum peak, the distributions are similar at different $r_\mathrm{cut}$ values, and results in a-\ce{Ga2O3} and a-\ce{Al2O3} are comparable with all chosen $r_\mathrm{cut}$ values. Figure~\ref{fig:d2min}(d, e) presents the visualization of a cross-section taken along the long edge. A slab with 5 \r A in thickness is visualized to avoid badly overlap of atoms. Each atom is individually colored by the momentary \dtmin\ value in accordance with the background color in Figure~\ref{fig:d2min}(a-c).

At $r_\mathrm{cut}=4.6$ \r A, a-\ce{Al2O3} and a-\ce{Ga2O3} show very similar morphology with respect to the size of the localized deformation regions. Although a gradient color coding method in Figure~\ref{fig:d2min}(d) helps to illustrate the transition from static to deformed regions, it makes comparison between these two materials difficult. Because of the highly comparable \dtmin\ distribution shown in Figure~\ref{fig:d2min}(b, c), a constant threshold value is used for both materials to divide the atoms into low \dtmin\ and high \dtmin\ groups, colored in red and blue, respectively. This helps in a direct comparison of the high \dtmin\ region size between a-\ce{Ga2O3} and a-\ce{Al2O3}. For each subfigure in Figure~\ref{fig:d2min}(d-e), we see that when using a smaller cutoff, we see more but smaller regions in both systems. The larger cutoff, on the other hand, shows fewer but larger regions of plasticity. Comparison of the visualization results in these two materials shows a slight difference in the high \dtmin\ region size. To further investigate the plastic deformation ability of these two materials, a quantitative characterization of the whole structure is needed.

\begin{figure*}[ht!]
\centering\includegraphics[width=\linewidth]{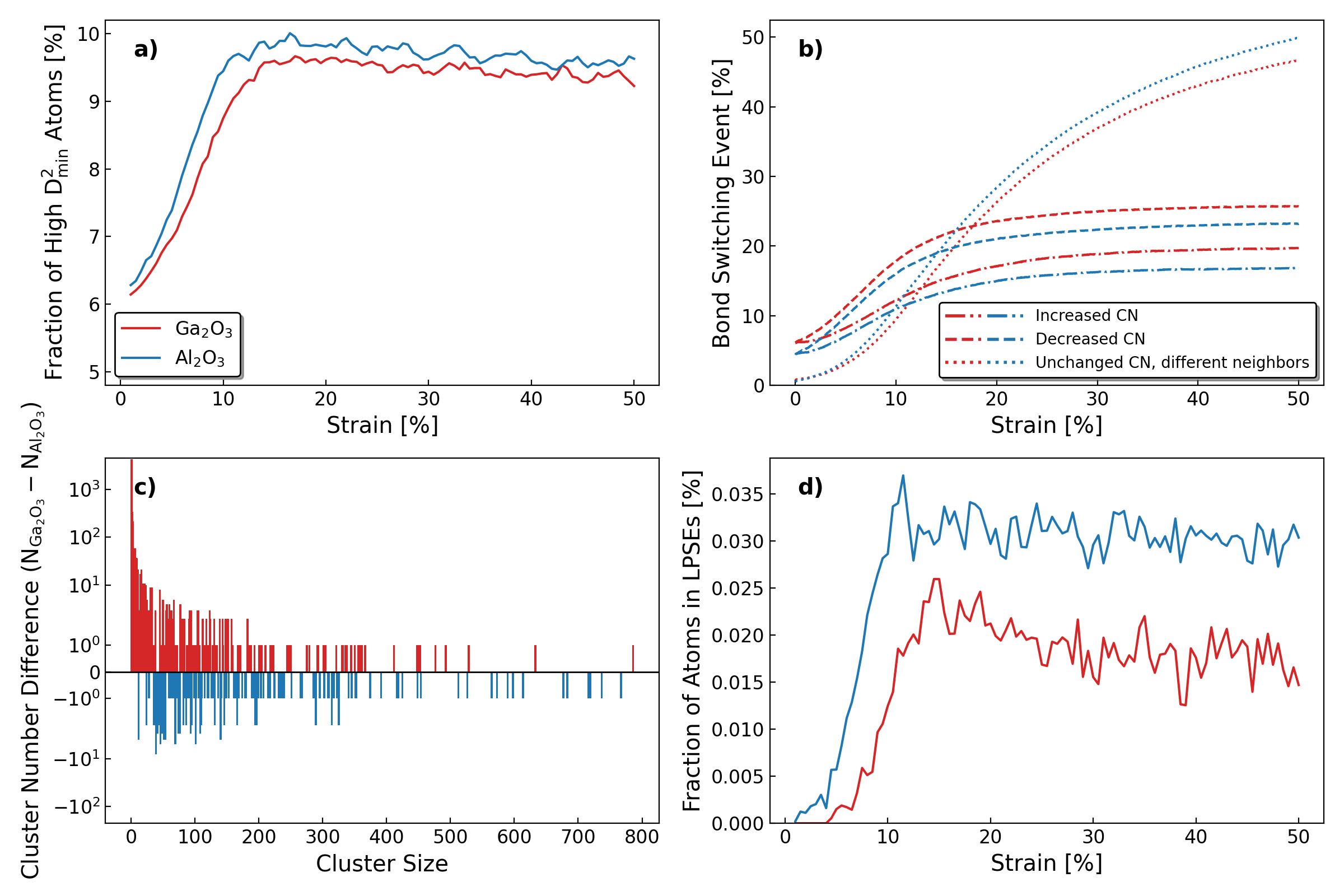}
\caption{(a) The fraction of high \dtmin\ atoms as a function of strain during the tensile test. The high \dtmin\ atom here means the atoms that have \dtmin\ value greater than two times the average \dtmin. (b) Bond change events compared to the unstrained structure. The results are classified as atoms with increased coordination numbers, decreased coordination numbers, unchanged coordination numbers but different bonded atoms. Absolute numbers are normalized according to the system size. (c) Difference in cluster number between a-\ce{Ga2O3} and a-\ce{Al2O3} at 50\% strain as a function of cluster size. (d) The fraction of atoms in LPSEs as a function of strain during tensile test. A LPSE is here defined as a high \dtmin\ atom cluster with more than 200 atoms.}
\label{fig:lpse}
\end{figure*}

As defined by Frankberg \textit{et al.}~\cite{Frankberg2019, Frankberg2023}, a LPSE is a highly deformed local volume of atoms and in simulations they can be captured by the momentary \dtmin\ characterization. The LPSEs play an important role during the plastic deformation of a-\ce{Al2O3} but has not yet been applied in the simulations of a-\ce{Ga2O3}. To obtain the fraction of atoms in LPSEs in a-\ce{Ga2O3}, the \dtmin\ results of the tensile test simulation are further analyzed. Although the average \dtmin\ values are close, they fluctuate during the tensile test simulation. Therefore, we use a threshold value of two times the average \dtmin\ at each moment to filter out high \dtmin\ atoms, as presented in Figure~\ref{fig:lpse}. We can see in both materials that the total fraction of high \dtmin\ atoms increases with strain and level approximately to 10\%, however in a-\ce{Al2O3} the material response to strain is faster leading to faster increase of the fraction of high \dtmin\ atoms and reaching a maximum peak of atoms associated with LPSE at a 2\% lower strain compared to a-\ce{Ga2O3}. The strain at which the fraction of high \dtmin\ atoms levels is already past the stress maximum, but matches the point when average momentary \dtmin\ levels in the two materials, as shown in Figure~\ref{fig:stress}(a). This shows that the stress is associated with the rate of atoms shifting to high \dtmin\ values and not with the absolute amount of them. Therefore, as the atoms associated with LPSEs are increasing in the system, the rate that the stress is increasing gradually lowers until it levels and finally obtains a negative slope. Next, the negative stress slope reverts towards a level slope when reaching equilibrium conditions and steady state flow. However, a slight negative slope in stress is observed in both materials up to 50\% strain indicating a gradual and loss of material strength under tensile strain. The similar fraction of high \dtmin\ atoms in both a-\ce{Ga2O3} and a-\ce{Al2O3} indicate that momentarily they have a very similar total amount of atoms involved in a comparable degree of localized deformation. Figure~\ref{fig:lpse}(b) presents the accumulation of bond switching events in these two materials. As a typical characterization of short-range ordering, it again shows slight differences during the tensile tests. Compared to the unstrained structure, the atom fractions with increased and decreased coordination number (CN) have initial values of 4\% for a-\ce{Al2O3} and 6\% for a-\ce{Ga2O3}, respectively. The fluctuation of bond lengths around the cut-off distance result in equal fraction of increased and decreased CN. As strain increases, the fractions of increased and decreased CN first increase in the two systems, then slow down and level at approximately 20\% and 26\% for a-\ce{Ga2O3}, 17\% and 23\% for a-\ce{Al2O3}, respectively. The fraction of decreased CN is greater than that of increased CN for both materials, indicating a slight but the same degree of expansion of the systems' volume. The fractions of unchanged CN with different neighbors starts from 0 at and quickly increase to 47\% and 50\% in a-\ce{Ga2O3} and a-\ce{Al2O3} at the end of the tensile tests, respectively. Therefore, majority of the atom translocations occur in both materials by changing neighboring atoms while retaining the local envirionment. Compared to Figure~\ref{fig:lpse}(a-b), histogram of high \dtmin\ atom clusters' size difference at 50\% strain presented in Figure~\ref{fig:lpse}(c) gives us some evident difference between a-\ce{Ga2O3} and a-\ce{Al2O3} finally. We see that a-\ce{Ga2O3} has significantly more small clusters, \textit{i.e.}, less than 20 atoms. While for a-\ce{Al2O3}, it has a much greater number of clusters in the range of 500 to 800 atoms. This clearly shows the spatial distribution difference in these systems. Then quantitatively, results of the fraction of atoms associated with LPSEs presented in Figure~\ref{fig:lpse} show evident differences between these two materials. Notably, a-\ce{Al2O3} has approximately 60\% higher fraction of atoms in LPSEs than a-\ce{Ga2O3}. The results indicate that although the two materials and structures have comparable amounts of highly plastic atoms, in a-\ce{Al2O3} the LPSE regions are larger and can react to applied stress faster.

\section{Discussion}

A comprehensive computational study is conducted in this work to first verify the reliability of the ML-IAP developed for generalized modeling of \ce{Ga2O3}~\cite{Zhao2023}, focusing on the amorphous state of this material. The computational results show good consistency with existing experimental results, such as density~\cite{Dingwell1992, Yu2003, Han2020, liu2023}, and mechanical properties~\cite{battu2018}. Comparisons are also performed with DFT data, but we note that the preparation conditions are critical for the amorphous structure, and thus we should be careful when doing such comparisons. The crystallization temperature of a-\ce{Ga2O3} has also been reported experimentally~\cite{nagarajan_chemically_2008, Kumar2013, battu2018}, but no crystallization was observed at the cooling rates investigated in this study. In summary, in addition to the crystalline phases validated earlier~\cite{Zhao2023}, we show that the developed ML-IAP is efficient and accurate in describing the amorphous state of \ce{Ga2O3} as well. Oxide glasses typically exhibit low fracture toughness leading to flaw sensitivity during mechanical loading. Our simulations confirm that melt-quenching produces an a-\ce{Ga2O3} structure sufficiently free of intrinsic geometrical defects that prevents a fracture from nucleating and propagating.

In accordance with the striking chemical similarity of both a-\ce{Ga2O3} and a-\ce{Al2O3}, our simulations predict both materials to be poor and fragile glass formers. This explains why their pure amorphous phases are difficult to synthesize with current technology and therefore these oxides are mainly used as glass modifier components in glass engineering. Due to the evident similarities between a-\ce{Ga2O3} and a-\ce{Al2O3} and the exceptional room temperature plasticity having been reported in the latter, comparisons were made between these two materials to understand the plastic deforming ability of a-\ce{Ga2O3} in detail.  

Amorphous \ce{Ga2O3} shows a significantly higher glass transition temperature compared to a-\ce{Al2O3}. As shown earlier for a-\ce{Al2O3}~\cite{Zhang2024}, the glass transition of a-\ce{Ga2O3} is dependent on the cooling rate and will dictate the mechanical properties of the obtained material. Therefore, using the same quench rate allows a better comparison between the intrinsic material properties. At a commonly shared quench rate [Figure~\ref{fig:stress}(a)], \ce{Ga2O3} glass has a slightly lower yield stress than \ce{Al2O3} glass. After a high plastic strain at 50\%, a-\ce{Al2O3} is known to undergo healing towards an energetically favorable flow structure and converge towards the minimum flow stress~\cite{Zhang2024}. At 50\% strain a-\ce{Ga2O3} again shows a slightly lower flow stress and a slightly higher tendency to continue softening as a function of strain which indicates minor changes occurring in the glassy structure up to high strains. Neither of these materials undergo hardening under plastic strain which is typical for amorphous materials due to lack of persisting plasticity mediators, such as dislocations. With currently available technology, glass transition is difficult to verify by experiments in such poor glass formers given the ultra-high quench rates involved and therefore this value is reported here also as a future reference. 

The results show that a-\ce{Ga2O3} can withstand high plastic strain comparable to that characterized for a-\ce{Al2O3}, despite the interatomic potentials for the two materials being mathematically completely different (fully machine-learned versus analytical pair potential). This also serves as good cross-validation that the amorphous structures prepared for both materials are likely realistic. Building the link between structural and mechanical properties was based on one decisive observation that the same $r_\mathrm{cut}$ can be used for both materials. However, although the RDFs of these two materials justify the use of this $r_\mathrm{cut}$, we should still consider the approximately 8\% difference in the peak positions of the cation-oxygen pair distribution indicating a significantly longer bond length for Ga-O in comparison to Al-O. In principle, such a bond length difference can lead to an atomic density difference of approximately 26\%, which is close to the difference in simulated atomic densities of this work. The bond length additionally explains why the lower atomic density still does not lead to a lower average coordination number, as illustrated by the coordination distributions and polyhedral analysis results. As such structural differences will ultimately dictate the mechanical behavior, a more detailed analysis should be conducted in the future.

Although similar plastic deforming ability is observed in the macroscopic stress-strain behavior, the LPSE analysis shows distinct differences between these two oxide glasses. As a short review of how the LPSEs can reflect on the viscous creep plasticity, in Ref.~\cite{Frankberg2023}, it has been previously found that the number of atoms associated with the LPSEs in a-\ce{Al2O3} remain nearly constant with increasing strain rate. At high strain rates, this leads to localization of the available LPSEs to shear bands at the location of the shear stress maximum. Accordingly, in a-\ce{Al2O3} micropillar compression experiments at relatively high strain rates up to 10$^3$ s$^{-1}$, micropillar deformation proceeds dominantly by a slip-like shear band propagation. At low quasi-static strain rates the deformation becomes homogeneous in the whole volume with only a minor contribution of shear bands to the plasticity. Amorphous \ce{Ga2O3} shows a similar overall amount of high \dtmin\ atoms with a-\ce{Al2O3}, however the LPSEs are smaller with fewer atoms included. This could indicate that the LPSE nucleation rate is higher in a-\ce{Ga2O3}. The nucleation rate of LPSEs was found to be a critical mediator of plasticity in oxide glasses~\cite{Frankberg2023}. Therefore, a possible prediction regarding the mechanical behavior of a-\ce{Ga2O3} is that due to the higher LPSE nucleation rate, a-\ce{Ga2O3} is predicted to yield a more homogeneous deformation at varying strain rates and to resist shear band formation more compared to a-\ce{Al2O3}. This can possibly increase the damage tolerance of a-\ce{Ga2O3} as catastrophic shear band propagation is a known failure mechanism in amorphous materials. However, experimental verification of these predictions remain essential.

\section{Conclusions}

In summary, we investigate the room temperature plasticity of a-\ce{Ga2O3} with large-scale atomistic simulations based on a newly developed ML-IAP, tabGAP, and the results are compared with existing experimental and computational results. On an atomistic simulation time scale, the IAP can produce an a-\ce{Ga2O3} structure that has density and structural properties comparable to the reported properties of the material. Therefore, the new tabGAP IAP is efficient and reliable in modeling a-\ce{Ga2O3}. Computational results validate that following a melt-quenching preparation process, a glass transition occurs in \ce{Ga2O3}.

Tensile test simulation shows that overall a-\ce{Ga2O3} exhibits room temperature plasticity comparable to that earlier observed in a-\ce{Al2O3} with similar number of atoms momentarily exhibiting a high local plasticity. However, differences were also found. When deformed at the same strain rate, the two materials have different fractions of atoms in the LPSE clusters that mediate the plasticity. This indicates a higher LPSE nucleation rate for a-\ce{Ga2O3} which can increase the resistance of the material to shear banding, which is a known failure mechanism in amorphous materials. The results of this work show that the ML-IAP is a useful tool in the study of the mechanical properties of amorphous materials, providing predictive information for experimental study.

\section{Acknowledgments}

We acknowledge funding from the Academy of Finland project numbers 315451, 315453, 326426, 338750 and 332347.
J. Zhao acknowledges the National Natural Science Foundation of China under Grant 62304097; Guangdong Basic and Applied Basic Research Foundation under Grant 2023A1515012048; Shenzhen Fundamental Research Program under Grant JCYJ20230807093609019.
J. Byggmästar acknowledges funding from the Research Council of Finland Flagship programme: Finnish Center for Artificial Intelligence (FCAI).
The computational resources granted by the CSC – IT Center for Science projects 2003839 (LAPLAS Glass Plasticity at Room Temperature) and hy3898, Finland, and by the Finnish Grid and Cloud Infrastructure project (FGCI; urn:nbn:fi:research-infras-2016072533) are gratefully acknowledged.



\bibliographystyle{elsarticle-num} 

\bibliography{ref.bib}

\begin{thebibliography}{10}
\expandafter\ifx\csname url\endcsname\relax
  \def\url#1{\texttt{#1}}\fi
\expandafter\ifx\csname urlprefix\endcsname\relax\def\urlprefix{URL }\fi
\expandafter\ifx\csname href\endcsname\relax
  \def\href#1#2{#2} \def\path#1{#1}\fi

\bibitem{idota1997}
Y.~Idota, T.~Kubota, A.~Matsufuji, Y.~Maekawa, T.~Miyasaka, {Tin-based
  amorphous oxide: A high-capacity lithium-ion-storage material}, Science
  276~(5317) (1997) 1395--1397.
\newblock \href {https://doi.org/10.1126/science.276.5317.1395}
  {\path{doi:10.1126/science.276.5317.1395}}.

\bibitem{Kim2006}
J.~Y. Kim, S.~H. Kim, H.~H. Lee, K.~Lee, W.~Ma, X.~Gong, A.~J. Heeger, {New
  architecture for high-efficiency polymer photovoltaic cells using
  solution-based titanium oxide as an optical spacer}, Advanced Materials
  18~(5) (2006) 572--576.
\newblock \href {https://doi.org/10.1002/adma.200501825}
  {\path{doi:10.1002/adma.200501825}}.

\bibitem{Yuan2014}
F.~Yuan, L.~Huang, {Brittle to ductile transition in densified silica glass},
  Scientific Reports 4 (2014) 1--8.
\newblock \href {https://doi.org/10.1038/srep05035}
  {\path{doi:10.1038/srep05035}}.

\bibitem{Lane2015}
J.~M.~D. Lane, {Cooling rate and stress relaxation in silica melts and glasses
  via microsecond molecular dynamics}, Physical Review E 92~(1) (2015) 012320.
\newblock \href {https://doi.org/10.1103/PhysRevE.92.012320}
  {\path{doi:10.1103/PhysRevE.92.012320}}.

\bibitem{Frankberg2019}
E.~J. Frankberg, J.~Kalikka, F.~G. Ferré, L.~Joly-Pottuz, T.~Salminen,
  J.~Hintikka, M.~Hokka, S.~Koneti, T.~Douillard, B.~L. Saint, P.~Kreiml, M.~J.
  Cordill, T.~Epicier, D.~Stauffer, M.~Vanazzi, L.~Roiban, J.~Akola, F.~D.
  Fonzo, E.~Levänen, K.~Masenelli-Varlot, Highly ductile amorphous oxide at
  room temperature and high strain rate, Science 366 (2019) 864--869.
\newblock \href {https://doi.org/10.1126/science.aav1254}
  {\path{doi:10.1126/science.aav1254}}.

\bibitem{Frankberg2023}
E.~J. Frankberg, A.~Lambai, J.~Zhang, J.~Kalikka, S.~Khakalo, B.~Paladino,
  M.~Cabrioli, N.~G. Mathews, T.~Salminen, M.~Hokka, J.~Akola, A.~Kuronen,
  E.~Levänen, F.~D. Fonzo, G.~Mohanty, Exceptional microscale plasticity in
  amorphous aluminum oxide at room temperature, Advanced Materials (7 2023).
\newblock \href {https://doi.org/10.1002/adma.202303142}
  {\path{doi:10.1002/adma.202303142}}.

\bibitem{Phillips1981}
J.~C. Phillips, Topology of covalent non-crystalline solids {II}: Medium-range
  order in chalcogenide alloys and a-si(ge), Journal of Non-Crystalline Solids
  43 (1981) 37--77.
\newblock \href {https://doi.org/10.1016/0022-3093(81)90172-1}
  {\path{doi:10.1016/0022-3093(81)90172-1}}.

\bibitem{Thorpe1983}
M.~F. Thorpe, Continuous deformations in random networks, Journal of
  Non-Crystalline Solids 57 (1983) 355--370.
\newblock \href {https://doi.org/10.1016/0022-3093(83)90424-6}
  {\path{doi:10.1016/0022-3093(83)90424-6}}.

\bibitem{Falk1998}
M.~L. Falk, J.~S. Langer, Dynamics of viscoplastic deformation in amorphous
  solids, Physical Review E 57 (1998) 14.

\bibitem{Gaskell1997}
P.~H. Gaskell, Structure and properties of glasses — how far do we need to
  go?, Journal of Non-Crystalline Solids 222 (1997) 1--12.
\newblock \href {https://doi.org/10.1016/S0022-3093(97)90091-0}
  {\path{doi:10.1016/S0022-3093(97)90091-0}}.

\bibitem{Du2004}
J.~Du, A.~N. Cormack, The medium range structure of sodium silicate glasses: A
  molecular dynamics simulation, Journal of Non-Crystalline Solids 349 (2004)
  66--79.
\newblock \href {https://doi.org/10.1016/j.jnoncrysol.2004.08.264}
  {\path{doi:10.1016/j.jnoncrysol.2004.08.264}}.

\bibitem{Zhang2023}
J.~Zhang, E.~J. Frankberg, J.~Kalikka, A.~Kuronen, Room temperature plasticity
  in amorphous {SiO$_2$} and amorphous {Al$_2$O$_3$}: A computational and
  topological study, Acta Materialia 259 (2023) 119223.
\newblock \href {https://doi.org/10.1016/j.actamat.2023.119223}
  {\path{doi:10.1016/j.actamat.2023.119223}}.

\bibitem{Levin1998}
I.~Levin, D.~Brandon, Metastable alumina polymorphs: Crystal structures and
  transition sequences, Journal of the American Ceramic Society 81 (1998)
  1995--2012.
\newblock \href {https://doi.org/10.1111/j.1151-2916.1998.tb02581.x}
  {\path{doi:10.1111/j.1151-2916.1998.tb02581.x}}.

\bibitem{HILL1952}
V.~G. HILL, R.~ROY, E.~F. OSBORN, The system alumina‐gallia‐water, Journal
  of the American Ceramic Society 35 (1952) 135--142.
\newblock \href {https://doi.org/10.1111/j.1151-2916.1952.tb13087.x}
  {\path{doi:10.1111/j.1151-2916.1952.tb13087.x}}.

\bibitem{wang_band_2018}
T.~Wang, W.~Li, C.~Ni, A.~Janotti, Band gap and band offset of {Ga$_2$O$_3$}
  and {(Al$_x$Ga$_{1-x}$)$_2$O$_3$} alloys, Physical Review Applied 10~(1)
  (2018) 011003.
\newblock \href {https://doi.org/10.1103/PhysRevApplied.10.011003}
  {\path{doi:10.1103/PhysRevApplied.10.011003}}.

\bibitem{Han2020}
Z.~Han, H.~Liang, W.~Huo, X.~Zhu, X.~Du, Z.~Mei, Boosted {UV} photodetection
  performance in chemically etched amorphous {Ga$_2$O$_3$} thin‐film
  transistors, Advanced Optical Materials 8 (4 2020).
\newblock \href {https://doi.org/10.1002/adom.201901833}
  {\path{doi:10.1002/adom.201901833}}.

\bibitem{Mu2022}
S.~Mu, C.~G.~V. de~Walle, Phase stability of {(Al$_x$Ga$_{1-x}$)$_2$O$_3$}
  polymorphs: A first-principles study, Physical Review Materials 6 (2022)
  104601.
\newblock \href {https://doi.org/10.1103/PhysRevMaterials.6.104601}
  {\path{doi:10.1103/PhysRevMaterials.6.104601}}.

\bibitem{Tadjer2022}
M.~J. Tadjer, Toward gallium oxide power electronics ultrawide-bandgap
  semiconductors show promise for high-power transistors, Science 378 (2022)
  724--725.
\newblock \href {https://doi.org/10.1126/science.add2713}
  {\path{doi:10.1126/science.add2713}}.

\bibitem{Cui2017}
S.~Cui, Z.~Mei, Y.~Zhang, H.~Liang, X.~Du, Room-temperature fabricated
  amorphous {Ga$_2$O$_3$} high-response-speed solar-blind photodetector on
  rigid and flexible substrates, Advanced Optical Materials 5 (10 2017).
\newblock \href {https://doi.org/10.1002/adom.201700454}
  {\path{doi:10.1002/adom.201700454}}.

\bibitem{Liang2019}
H.~Liang, S.~Cui, R.~Su, P.~Guan, Y.~He, L.~Yang, L.~Chen, Y.~Zhang, Z.~Mei,
  X.~Du, Flexible {X}-ray detectors based on amorphous {Ga$_2$O$_3$} thin
  films, ACS Photonics 6 (2019) 351--359.
\newblock \href {https://doi.org/10.1021/acsphotonics.8b00769}
  {\path{doi:10.1021/acsphotonics.8b00769}}.

\bibitem{Qin2019}
Y.~Qin, S.~Long, Q.~He, H.~Dong, G.~Jian, Y.~Zhang, X.~Hou, P.~Tan, Z.~Zhang,
  Y.~Lu, C.~Shan, J.~Wang, W.~Hu, H.~Lv, Q.~Liu, M.~Liu, Amorphous gallium
  oxide-based gate-tunable high-performance thin film phototransistor for
  solar-blind imaging, Advanced Electronic Materials 5 (7 2019).
\newblock \href {https://doi.org/10.1002/aelm.201900389}
  {\path{doi:10.1002/aelm.201900389}}.

\bibitem{Schubert2016}
M.~Schubert, R.~Korlacki, S.~Knight, T.~Hofmann, S.~Schöche, V.~Darakchieva,
  E.~Janzén, B.~Monemar, D.~Gogova, Q.~T. Thieu, R.~Togashi, H.~Murakami,
  Y.~Kumagai, K.~Goto, A.~Kuramata, S.~Yamakoshi, M.~Higashiwaki, Anisotropy,
  phonon modes, and free charge carrier parameters in monoclinic
  $\beta$-gallium oxide single crystals, Physical Review B 93 (3 2016).
\newblock \href {https://doi.org/10.1103/PhysRevB.93.125209}
  {\path{doi:10.1103/PhysRevB.93.125209}}.

\bibitem{Furthmuller2016}
J.~Furthm{\"u}ller, F.~Bechstedt, Quasiparticle bands and spectra of
  {Ga$_2$O$_3$} polymorphs, Physical Review B 93 (3 2016).
\newblock \href {https://doi.org/10.1103/PhysRevB.93.115204}
  {\path{doi:10.1103/PhysRevB.93.115204}}.

\bibitem{ponce2020}
S.~Ponc{\'e}, F.~Giustino, Structural, electronic, elastic, power, and
  transport properties of {$\beta$-Ga$_2$O$_3$} from first principles, Physical
  Review Research 2 (7 2020).
\newblock \href {https://doi.org/10.1103/PhysRevResearch.2.033102}
  {\path{doi:10.1103/PhysRevResearch.2.033102}}.

\bibitem{Lion2022}
K.~Lion, P.~Pavone, C.~Draxl, Elastic stability of {Ga$_2$O$_3$}: Addressing
  the $\beta$-to-$\alpha$ phase transition from first principles, Physical
  Review Materials 6 (2022) 013601.
\newblock \href {https://doi.org/10.1103/PhysRevMaterials.6.013601}
  {\path{doi:10.1103/PhysRevMaterials.6.013601}}.

\bibitem{Zhang2020}
Z.~Zhang, S.~Ispas, W.~Kob, The critical role of the interaction potential and
  simulation protocol for the structural and mechanical properties of
  sodosilicate glasses, Journal of Non-Crystalline Solids 532 (3 2020).
\newblock \href {https://doi.org/10.1016/j.jnoncrysol.2020.119895}
  {\path{doi:10.1016/j.jnoncrysol.2020.119895}}.

\bibitem{tang_brittle--ductile_2023}
L.~Tang, M.~M. Smedskjaer, M.~Bauchy, The brittle-to-ductile transition in
  aluminosilicate glasses is driven by topological and dynamical heterogeneity,
  Acta Materialia 247 (2023) 118740.
\newblock \href {https://doi.org/10.1016/j.actamat.2023.118740}
  {\path{doi:10.1016/j.actamat.2023.118740}}.

\bibitem{Wu2021}
Y.~Wu, D.~Cao, Y.~Yao, G.~Zhang, J.~Wang, L.~Liu, F.~Li, H.~Fan, X.~Liu,
  H.~Wang, X.~Wang, H.~Zhu, S.~Jiang, P.~Kontis, D.~Raabe, B.~Gault, Z.~Lu,
  Substantially enhanced plasticity of bulk metallic glasses by densifying
  local atomic packing, Nature Communications 12 (12 2021).
\newblock \href {https://doi.org/10.1038/s41467-021-26858-9}
  {\path{doi:10.1038/s41467-021-26858-9}}.

\bibitem{Unke2021}
O.~T. Unke, S.~Chmiela, H.~E. Sauceda, M.~Gastegger, I.~Poltavsky, K.~T.
  Schütt, A.~Tkatchenko, K.-R. Müller, Machine learning force fields,
  Chemical Reviews 121 (2021) 10142--10186.
\newblock \href {https://doi.org/10.1021/acs.chemrev.0c01111}
  {\path{doi:10.1021/acs.chemrev.0c01111}}.

\bibitem{Liu2020}
Y.~B. Liu, J.~Y. Yang, G.~M. Xin, L.~H. Liu, G.~Csányi, B.~Y. Cao, Machine
  learning interatomic potential developed for molecular simulations on thermal
  properties of $\beta$-{Ga$_2$O$_3$}, Journal of Chemical Physics 153 (10
  2020).
\newblock \href {https://doi.org/10.1063/5.0027643}
  {\path{doi:10.1063/5.0027643}}.

\bibitem{liu2023}
Y.~Liu, H.~Liang, L.~Yang, G.~Yang, H.~Yang, S.~Song, Z.~Mei, G.~Cs{\'a}nyi,
  B.~Cao, Unraveling thermal transport correlated with atomistic structures in
  amorphous gallium oxide via machine learning combined with experiments,
  Advanced Materials (2023) 2210873\href
  {https://doi.org/10.1002/adma.202210873} {\path{doi:10.1002/adma.202210873}}.

\bibitem{Zhao2023}
J.~Zhao, J.~Byggmästar, H.~He, K.~Nordlund, F.~Djurabekova, M.~Hua, Complex
  {Ga$_2$O$_3$} polymorphs explored by accurate and general-purpose
  machine-learning interatomic potentials, npj Computational Materials 9 (2023)
  159.
\newblock \href {https://doi.org/10.1038/s41524-023-01117-1}
  {\path{doi:10.1038/s41524-023-01117-1}}.

\bibitem{hoover1985}
W.~G. Hoover, Canonical dynamics: Equilibrium phase-space distributions,
  Physical Review A 31 (1985) 1695--1697.
\newblock \href {https://doi.org/10.1103/PhysRevA.31.1695}
  {\path{doi:10.1103/PhysRevA.31.1695}}.

\bibitem{thompson_lammps_2022}
A.~P. Thompson, H.~M. Aktulga, R.~Berger, D.~S. Bolintineanu, W.~M. Brown,
  P.~S. Crozier, P.~J. {in't Veld}, A.~Kohlmeyer, S.~G. Moore, T.~D. Nguyen,
  R.~Shan, M.~J. Stevens, J.~Tranchida, C.~Trott, S.~J. Plimpton, {{LAMMPS}} -
  a flexible simulation tool for particle-based materials modeling at the
  atomic, meso, and continuum scales, Computer Physics Communications 271
  (2022) 108171.
\newblock \href {https://doi.org/10.1016/j.cpc.2021.108171}
  {\path{doi:10.1016/j.cpc.2021.108171}}.

\bibitem{Gutierrez2002}
G.~Guti{\'{e}}rrez, B.~Johansson, {Molecular dynamics study of structural
  properties of amorphous Al$_2$O$_3$}, Physical Review B 65~(10) (2002)
  104202.
\newblock \href {https://doi.org/10.1103/PhysRevB.65.104202}
  {\path{doi:10.1103/PhysRevB.65.104202}}.

\bibitem{Matsui1994}
M.~Matsui, {A transferable interatomic potential model for crystals and melts
  in the system CaO-MgO-Al$_2$O$_3$-SiO$_2$}, Mineralogical Magazine 58A~(2)
  (1994) 571--572.
\newblock \href {https://doi.org/10.1180/minmag.1994.58A.2.34}
  {\path{doi:10.1180/minmag.1994.58A.2.34}}.

\bibitem{Ovito}
A.~Stukowski, {Visualization and analysis of atomistic simulation data with
  {OVITO}-the open visualization tool}, {Modelling and Simulation in Materials
  Science and Engineering} {18}~({1}) (2010).
\newblock \href {https://doi.org/{10.1088/0965-0393/18/1/015012}}
  {\path{doi:{10.1088/0965-0393/18/1/015012}}}.

\bibitem{Yu2003}
Z.~Yu, C.~D. Overgaard, R.~Droopad, M.~Passlack, J.~K. Abrokwah, Growth and
  physical properties of {Ga$_2$O$_3$} thin films on gaas (001) substrate by
  molecular-beam epitaxy, Applied Physics Letters 82 (2003) 2978--2980.
\newblock \href {https://doi.org/10.1063/1.1572478}
  {\path{doi:10.1063/1.1572478}}.

\bibitem{Dingwell1992}
D.~B. Dingwell, Density of {Ga$_2$O$_3$} liquid, Journal of the American
  Ceramic Society 75 (1992) 1656--1657.
\newblock \href {https://doi.org/10.1111/j.1151-2916.1992.tb04239.x}
  {\path{doi:10.1111/j.1151-2916.1992.tb04239.x}}.

\bibitem{Zhang2024}
J.~Zhang, M.~D. Meulder, E.~J. Frankberg, A.~Kuronen, Dependence between glass
  transition and plasticity in amorphous aluminum oxide: A molecular dynamics
  study, Journal of Non-Crystalline Solids 628 (2024) 122840.
\newblock \href {https://doi.org/10.1016/j.jnoncrysol.2024.122840}
  {\path{doi:10.1016/j.jnoncrysol.2024.122840}}.

\bibitem{vogel_temperaturabhangigkeitsgesetz_1921}
H.~Vogel, Das {Temperaturabhangigkeitsgesetz} der {Viskositat} von
  {Flussigkeiten}, Physikalische Zeitschrift 22 (1921) 645--646.

\bibitem{fulcher_analysis_1925}
G.~S. Fulcher, {Analysis} {of} {recent} {measurements} {of} {the} {viscosity}
  {of} {glasses}, Journal of the American Ceramic Society 8~(6) (1925)
  339--355.
\newblock \href {https://doi.org/10.1111/j.1151-2916.1925.tb16731.x}
  {\path{doi:10.1111/j.1151-2916.1925.tb16731.x}}.

\bibitem{tammann_abhangigkeit_1926}
G.~Tammann, W.~Hesse, Die {Abhängigkeit} der {Viscosität} von der
  {Temperatur} bie unterkühlten {Flüssigkeiten}, Zeitschrift für
  anorganische und allgemeine Chemie 156~(1) (1926) 245--257.
\newblock \href {https://doi.org/10.1002/zaac.19261560121}
  {\path{doi:10.1002/zaac.19261560121}}.

\bibitem{Angell2000}
C.~A. Angell, K.~L. Ngai, G.~B. McKenna, P.~F. McMillan, S.~W. Martin,
  Relaxation in glassforming liquids and amorphous solids, Journal of Applied
  Physics 88 (2000) 3113--3157.
\newblock \href {https://doi.org/10.1063/1.1286035}
  {\path{doi:10.1063/1.1286035}}.

\bibitem{Mauro2009}
J.~C. Mauro, Y.~Yue, A.~J. Ellison, P.~K. Gupta, D.~C. Allan, Viscosity of
  glass-forming liquids, Proceedings of the National Academy of Sciences 106
  (2009) 19780--19784.
\newblock \href {https://doi.org/10.1073/pnas.0911705106}
  {\path{doi:10.1073/pnas.0911705106}}.

\bibitem{Yuan2012}
F.~Yuan, L.~Huang, Molecular dynamics simulation of amorphous silica under
  uniaxial tension: From bulk to nanowire, Journal of Non-Crystalline Solids
  358 (2012) 3481--3487.
\newblock \href {https://doi.org/10.1016/j.jnoncrysol.2012.05.045}
  {\path{doi:10.1016/j.jnoncrysol.2012.05.045}}.

\bibitem{SFzhao2024crystallization}
J.~Zhao, J.~Garc{\'i}a~Fern{\'a}ndez, A.~Azarov, R.~He, {\O}.~Prytz,
  K.~Nordlund, M.~Hua, F.~Djurabekova, A.~Kuznetsov, Crystallization instead of
  amorphization in collision cascades in gallium oxide, arXiv (2024).
\newblock \href {http://arxiv.org/abs/2401.07675} {\path{arXiv:2401.07675}},
  \href {https://doi.org/10.48550/arXiv.2401.07675}
  {\path{doi:10.48550/arXiv.2401.07675}}.

\bibitem{battu2018}
A.~K. Battu, C.~V. Ramana, Mechanical properties of nanocrystalline and
  amorphous gallium oxide thin films, Advanced Engineering Materials 20 (11
  2018).
\newblock \href {https://doi.org/10.1002/adem.201701033}
  {\path{doi:10.1002/adem.201701033}}.

\bibitem{Garcia2013}
F.~G. Ferr{\'e}, E.~Bertarelli, A.~Chiodoni, D.~Carnelli, D.~Gastaldi, P.~Vena,
  M.~G. Beghi, F.~D. Fonzo, The mechanical properties of a nanocrystalline
  {Al$_2$O$_3$/a-Al$_2$O$_3$} composite coating measured by nanoindentation and
  {Brillouin} spectroscopy, Acta Materialia 61 (2013) 2662--2670.
\newblock \href {https://doi.org/10.1016/j.actamat.2013.01.050}
  {\path{doi:10.1016/j.actamat.2013.01.050}}.

\bibitem{novikov_poissons_2004}
V.~N. Novikov, A.~P. Sokolov, Poisson's ratio and the fragility of
  glass-forming liquids, Nature 431~(7011) (2004) 961--963.
\newblock \href {https://doi.org/10.1038/nature02947}
  {\path{doi:10.1038/nature02947}}.

\bibitem{Wilding2010}
M.~C. Wilding, C.~J. Benmore, J.~K. Weber, High-energy {X}-ray diffraction from
  aluminosilicate liquids, Journal of Physical Chemistry B 114 (2010)
  5742--5746.
\newblock \href {https://doi.org/10.1021/jp907587e}
  {\path{doi:10.1021/jp907587e}}.

\bibitem{nagarajan_chemically_2008}
L.~Nagarajan, R.~A. De~Souza, D.~Samuelis, I.~Valov, A.~Börger, J.~Janek,
  K.-D. Becker, P.~C. Schmidt, M.~Martin, A chemically driven insulator–metal
  transition in non-stoichiometric and amorphous gallium oxide, Nature
  Materials 7~(5) (2008) 391--398.
\newblock \href {https://doi.org/10.1038/nmat2164}
  {\path{doi:10.1038/nmat2164}}.

\bibitem{Kumar2013}
S.~S. Kumar, E.~J. Rubio, M.~Noor-A-Alam, G.~Martinez, S.~Manandhar,
  V.~Shutthanandan, S.~Thevuthasan, C.~V. Ramana, Structure, morphology, and
  optical properties of amorphous and nanocrystalline gallium oxide thin films,
  Journal of Physical Chemistry C 117 (2013) 4194--4200.
\newblock \href {https://doi.org/10.1021/jp311300e}
  {\path{doi:10.1021/jp311300e}}.

\end{thebibliography}

% \include{sup.tex}



% \end{thebibliography}

%----------------------------------------------------------------------------------------------------------

% \appendix
% \setcounter{figure}{0}
% \renewcommand{\thefigure}{S\arabic{figure}}
%
% \section*{Supplementary Materials}

\end{document}